\documentclass[aps,twocolumn,prl,reprint,amsmath,amssymb,showpacs,superscriptaddress]{revtex4-1}

\usepackage{graphicx}
\usepackage{dcolumn}
\usepackage{bm}
\usepackage{amsmath}
\usepackage{color}
\usepackage{mathrsfs}
\usepackage{float}
\usepackage{txfonts}
\usepackage{wasysym}

\makeatletter

\newcommand{\Rmnum}[1]{\expandafter\@slowromancap\romannumeral #1@}
\makeatother
\begin{document}

\title{Tomography of zero-energy end modes in topological superconducting wires} 

\author{A. A. Aligia}
\affiliation{Centro At\'{o}mico Bariloche, Comisi\'{o}n Nacional
de Energ\'{\i}a At\'{o}mica, 8400 Bariloche, Argentina}
\affiliation{Instituto Balseiro, Comisi\'{o}n Nacional
de Energ\'{\i}a At\'{o}mica, 8400 Bariloche, Argentina}
\affiliation{Consejo Nacional de Investigaciones Cient\'{\i}ficas y T\'ecnicas,
1025 CABA, Argentina}

\author{D. P\'erez Daroca}
\affiliation{Gerencia de Investigaci\'on y Aplicaciones, Comisi\'on Nacional de
Energ\'ia At\'omica, 1650 San Mart\'{\i}n, Buenos Aires, Argentina}
\affiliation{Consejo Nacional de Investigaciones Cient\'{\i}ficas y T\'ecnicas,
1025 CABA, Argentina}

\author{Liliana Arrachea}
\affiliation{International Center for Advanced Studies, Escuela de Ciencia y Tecnolog\'{\i}a and ICIFI, 
Universidad Nacional de San Mart\'{\i}n,  25 de Mayo y Francia, 1650 Buenos Aires, Argentina}
\affiliation{Consejo Nacional de Investigaciones Cient\'{\i}ficas y T\'ecnicas,
1025 CABA, Argentina}

\begin{abstract}
We describe the Majorana zero modes in topological hybrid superconductor-semiconductor wires
 with spin-orbit coupling and magnetic field, in terms of generalized Bloch coordinates 
 $\varphi, \theta, \delta$.
When the spin-orbit coupling and the magnetic field are perpendicular,
$\varphi$ and $\delta$ are universal in an appropriate coordinate system.
We show how to extract  the angle $\theta$ from the behavior of the Josephson current-phase relation, which enables tomography of the Majorana modes.
Simple analytical expressions describe accurately the numerical results.
\end{abstract}

\date{\today}

\maketitle
Topological superconductors host Majorana 
zero modes (MZMs)  localized at the edges of the system\cite{kitaev-model,ali}. Detection and manipulation MZMs is motivated by their potential use  for implementing  topological quantum computation\cite{kitaev-qc,kitaev}.
Quantum wires with 
spin-orbit coupling (SOC), proximity-induced s-wave superconductivity and 
a magnetic field $\vec{B}$ having a component
perpendicular to the direction of the SOC \cite{wires1,wires2}, are one of the most prominent systems. 
Several 
works investigated 
 realizations of this platform for topological superconductivity in wires 
 of InAs \cite{wires-exp1,wires-exp2,wires-exp3,wires-exp4,wires-exp5,wires-jose}.

The existence of MZMs  leads to signatures in the behavior
of the Josephson current-phase relation (CPR). In the ac case, the periodicity as a function of
the phase bias $\phi $ is $4\pi $ for non-time reversal
invariant, \cite{kitaev-model,trs1,trs2,trs3,oro-jose,trs4,trs5,trs6,trs6m,trs7,trs8,trs9,trs10,trs11,trs12,trs13,trs14,trs15,schuray,san-jose,pikulin,
lee, bad, pen, sti-sau} 
and  time-reversal invariant  \cite{chung,ady,mellars,liu,nos,sch-fu,jorg,review-tritops,cata1,haim,alvarado,jason20} families, in contrast to the $2\pi $ one of the
ordinary superconductors.

In the topological superconducting phase of the quantum wires proposed in
Refs. \cite{wires1,wires2}, the zero modes have a non-trivial spin texture 
\cite{sticlet,klino,prada}. In contrast to what might be naively expected,
the spin
density from the exact solution of the model Hamiltonian shows that
the zero modes have magnetization
components perpendicular to $\vec{B}$ and the SOC axis.
Remarkably, for perpendicular $\vec{B}$ and SOC,
the components of the spin polarization perpendicular to $\vec{B}$  
are also perpendicular to the SOC and have opposite 
signs at the two ends of the wire \cite{wires1,sticlet}.

 We introduce a geometrical characterization of the MZMs in terms of their
generalized Bloch coordinates (GBC), i.e., Bloch coordinates $\varphi, \theta$ associated to the spin orientation and a phase $\delta$.
 We show that, when $\vec{B}$ and the SOC are perpendicular, there exist an {\em easy coordinate frame (ECF)} where  $\varphi$ and $\delta$ are universal and can be exactly calculated by  symmetry arguments, up to a sign that can be obtained from
 the solution in particular limits. We present a low-energy effective Hamiltonian to describe the MZMs in Josephson junctions and show that the angle $\theta$ can be inferred from the behavior of the CPR in suitable junctions, hence enabling a full tomography of these modes.

We consider a lattice version of the model for topological superconducting wires introduced in Refs. \cite{wires1,wires2}, with arbitrary orientations of $\vec{B}$ and SOC \cite{osca,rex}. 
The corresponding 
Hamiltonian is $H_{\rm w}=H_{0}+H_{\Delta }$, with 
\begin{eqnarray}
H_{0} &=&\sum_{\ell}{\bf c}_{\ell}^{\dagger }\left( -t\;\sigma_0-i\vec{\lambda}\cdot \vec{\sigma} \right)
{\bf c}_{\ell+1}+\text{H.c.}  \label{hamw} \\
&-&\sum_{\ell} {\bf c}_{\ell}^{\dagger }  \left( \vec{B} \cdot  \vec{\sigma}+\mu \sigma_0 \right)
{\bf c}_{\ell},\;\;\;\;H_{\Delta }=\Delta \sum_{\ell}c_{ \ell \uparrow
}^{\dagger }c_{\ell \downarrow }^{\dagger }+\text{H.c.},  \notag
\end{eqnarray}
where $\ell$ labels sites of a 1D lattice and ${\bf c}_{\ell}=(c_{\ell \uparrow}, c_{\ell \downarrow})^T$.
$\vec{B}=B \vec{n}_B$ and $\vec{\lambda}=\lambda  \vec{n}_{\lambda }$, with $B, \lambda \geq 0$, 
are the magnetic field and the SOC oriented 
along the spacial directions $\vec{n}_B$ and 
 $\vec{n}_{\lambda }$, respectively.
 The components of the vector $\vec{\sigma}=\left( \sigma _{x},
\sigma_{y},\sigma _{z}\right) $ are the Pauli matrices and $\sigma_0$ is the 2$\times$2 unitary matrix.
This model has a
topological phase 
provided that $\vec{n}_{\lambda }$ and  $\vec{n}_B$ are not parallel. 
The evaluation of topological invariants \cite{tewa,budich},
leads to the following expressions for the boundaries  
\begin{equation}
|2|t|-r| <|\mu| < |2|t|+r|, \;\;\;\;\;\;\;\;\; B \; |\vec{n}_{\lambda} \cdot \vec{n}_B|<|\Delta| <B, \label{bound}
\end{equation}
with $r =\sqrt{B^{2}-\Delta^{2}}$.

The MZMs of the Hamiltonian of Eq. (\ref{hamw}) 
can be written as $\eta_{\nu } =\gamma{_\nu }^{\dagger }+\gamma{_\nu }$,
where $\nu =L, R$
denotes the left and right end of the wires, respectively. 
We assume that the spin of 
$\gamma_{\nu }^{\dagger }$ 
is oriented along the Bloch vector
$\vec{n}_{\nu }=\left( \cos \theta _{\nu }\sin \varphi _{\nu },\cos \theta_{\nu }\cos \varphi _{\nu },\sin \theta _{\nu }\right) $. 
The angles $\theta_{\nu}$ and $\varphi_{\nu}$ in the Bloch sphere, as well as a phase 
$\delta _{\nu }$ --- defined $\mbox{mod}(\pi)$ --- are the GBC, which fully characterize the  MZM through
\begin{eqnarray}
\gamma_{\nu }^{\dagger }=e^{i\delta _{\nu }}\left[ \cos (\theta _{\nu }/2)c_{\nu
\uparrow }^{\dagger }+e^{i\varphi _{\nu }}\sin (\theta _{\nu }/2)c_{\nu
\downarrow }^{\dagger }\right].  \label{majo}
\end{eqnarray}
Here,  $c_{\nu s}^{\dagger }$ are fermionic  creation
operators associated to the basis of $H_{\rm w}$, acting at the ends of the wire (usually including a few sites).
Importantly, not only 
the angles $\theta_{\nu}$ and $\varphi_{\nu}$, but also 
$\delta_{\nu}$
depend on the choice of the reference frame. 
The change in $\vec{n}_{\nu}$ under a rotation of the coordinate system is
a routine exercise. The corresponding change of $\delta_{\nu}$ leads to a function $\xi_{L,R} \left( \vec{n}_L,\vec{n}_R \right)$, 
 --- see Eq. (S5) in the SM \cite{sm} --- which is a vector potential that depends on $\vec{n}_{\nu}$ but not on $\delta_{\nu}$, generated by a twist between the spin directions \cite{hamamoto}. 
 The quantity
\begin{equation}\label{rel}
\delta_{L,R}= \delta_{L} - \delta_{R}- \xi_{L,R}\left( \vec{n}_L,\vec{n}_R \right)  \; \mbox{mod}(\pi),
\end{equation}
 is invariant under rotations.  Notice that the SU(2) invariance of $\delta_{L,R}$ is expected since it appears in the evaluation of expectation values of observables, in particular, the current through the
 closing contact of a ring  formed with the wire, which is threaded by a magnetic flux.
  In addition, the scalar product of the two unit vectors, $\vec{n}_L \cdot \vec{n}_R$, is also an SU(2)-invariant.
  
We  notice that when
$\vec{n}_{\lambda }\cdot \vec{y} = \vec{n}_{B}\cdot \vec{y} = 0$, the Hamiltonian 
is invariant under inversion (defined by 
$\ell \leftrightarrow N+1-\ell$, 
for a chain with $N$ sites) and complex conjugation, implying
\begin{equation}
\delta_R=-\delta_L, \;\;\;\; \theta _R=\theta _L=\theta \;\;\;\;\varphi_R=-\varphi _L. \label{c1}
\end{equation}

The Hamiltonian  is also invariant under inversion
and simultaneous change in the
sign of $\lambda $.
For $\vec{n}_{\lambda }\cdot \vec{n}_{B}=0$ and $\vec{n}_B || \vec{z}$, the latter change of sign can
be absorbed in a gauge transformation 
$\tilde{c}_{\ell \uparrow }^{\dagger}=ic_{\ell \uparrow }^{\dagger }$, 
$\tilde{c}_{\ell \downarrow }^{\dagger}=-ic_{\ell \downarrow }^{\dagger }$. 
Therefore the MZM  for $\nu =R$, 
has the same form as the one for $\nu =L$, replacing the operators $c_{\ell \sigma}^{\dagger }$ 
at the left end by the $\tilde{c}_{\ell \sigma}^{\dagger }$ at the
right. Hence, the GBC at the two ends are related as
\begin{equation}
\delta_R=\delta_L \pm \frac{\pi }{2}, \;\;\; \theta _R=\theta _L=\theta, \;\;\; 
\varphi_R=\varphi _L+\pi . \label{lr}
\end{equation}
This means that the Bloch vectors of  the MZMs have components perpendicular to $\vec{n}_B$ with opposite signs at the two edges, a conclusion that has been previously reached
after the  explicit calculation of the wave function in particular frames \cite{sticlet,prada}. 
We conclude that this property does not depend on the choice of the coordinate frame, 
since the relative tilt of the spin orientations is invariant under rotations.
Furthermore, combining  with  the condition of Eq. (\ref{c1}), we  identify an ECF: 
$\vec{n}_B || \vec{z}$ and $\vec{n}_{\lambda } || \vec{x}$. In that frame we have
\begin{equation}
\delta_R=-\delta_L = \pm \pi/4, \;\;\;\varphi_R=-\varphi _L = \pm \pi/2, \;\;\; \theta _R=\theta _L=\theta.
\label{inver}
\end{equation}

To conclude the full characterization of the MZMs in this frame, 
we still need to define the signs in Eq. (\ref{inver})
and find the relation between $\theta_{\nu}$ and the
parameters of the Hamiltonian [Eq. (\ref{hamw})]. 
In what follows we present results for the case $B \gg \lambda, |\Delta|$, which 
by continuity leads to  the exact values of
$\varphi_{\nu}$ and $\delta_{\nu}$ in the full parameter space. 
In the 
SM \cite{sm}, we show that they coincide with the values for these parameters obtained from the calculation of the continuum version of the model in Ref. \cite{wires1} in the limit of dominant SOC 

The limit of dominant magnetic field is intuitively related to Kitaev's model, although in the present case, the MZMs are not fully polarized in the direction of $\vec{B}$, 
as explained before. 

Our aim now is to
explicitly calculate the GBC of the two MZMs
as functions of the Hamiltonian parameters $\lambda, \Delta, B, \mu$ when $B$ dominates,
 in the ECF.
To this end, it is useful to 
rewrite the Hamiltonian $H_{\rm w}$ of the wires in the basis that diagonalizes $H_{0}$ in Eq. (\ref{hamw}).  
We introduce the unitary
transformation in reciprocal space $d_{k+}=u_{k}c_{k\uparrow}+v_{k}c_{k\downarrow }$, $d_{k-}=-v_{k}c_{k\uparrow }+u_{k}c_{k\downarrow }$, 
being  $u_{k},\;v_k/\mbox{sgn} (\lambda _{k})=\sqrt{(1\pm B/r_{k})/2}$, 
  with $r_{k}=\sqrt{ \lambda _{k}^{2}+B^{2}}$, and $\lambda_{k}=2 \lambda \sin k$. This leads to
\begin{equation}
H_{\rm w} =\sum_{k,s=+,-}\left( \varepsilon _{k,s} \; d_{ks}^{\dagger }d_{ks}+ \Delta _{k}^{T}d_{ks}^{\dagger
}d_{-ks}^{\dagger }\right)+  \sum_{k} \Delta _{k}^{S} d_{k+}^{\dagger }d_{-k-}^{\dagger}+ \text{H.c.}
\label{hkt} 
\end{equation}
being $\varepsilon _{k,s}=\xi _{k}\mp r_{k}$ with $\xi _{k}=-2t\cos k-\mu $.
The pairing interaction contains a triplet component with p-wave symmetry
$\Delta_{k}^{T}=- \lambda_k \Delta /r_k$
 --- notice that $\lambda_k$ is
an odd function of $k$ ---
and a singlet one, $\Delta _{k}^{S}=
B \Delta/ r_k $. 

For $B\gg \Delta \gg \lambda$, the transformed model can be solved analytically 
with the method of Alase \textit{et al.} \cite{alas1,alas2,entangle,alcam}
(see SM \cite{sm}, for details). For
$t, \Delta >0$, and $\mu = -B, \Delta$, the results are
\begin{eqnarray}
\delta_L& =&- \delta_R=  \frac{ \pi}{4}, \;\;\;\;\; \varphi _L =-\varphi_R= - \frac{\pi}{2}, \label{apl} \\
\theta &\sim & \frac{\Delta }{B+\sqrt{(B^2-t^2)}}
+O(\frac{ \lambda}{B}), \;\;\;\;\;\;\; \vec{n}_{B } = \vec{z}, \;\;\vec{n}_{\lambda } = \vec{x}. \nonumber
\end{eqnarray}
While Eq. (\ref{inver}) gives the values of $\delta_{\nu}$ and 
$\varphi_{\nu}$ up to a sign, Eq. (\ref{apl}) gives their exact values. Although the calculation was done for dominant $\vec{B}$, this result is valid for continuity in the whole topological phase
with $B,t, \Delta, \lambda >0, \mu <0$. The corresponding values for the opposite signs of these parameters can be deduced by means of symmetry arguments \cite{syma}.
  
It is important to highlight that the previous results and appropriate SU(2) rotations
permit to obtain exactly $\delta_\nu$ and $\varphi_\nu$ in any coordinate system for any value of the parameters of Eq. (\ref{hamw}) with $\vec{n}_{\lambda} \cdot \vec{n}_B=0$, while 
$\theta$ needs an explicit calculation. Our goal now is to show that this angle can be inferred from the behavior of the CPR in suitable junctions.

To calculate the CPR  we consider two wires ${\rm w1}$ and ${\rm w2}$ 
 with different phases $\phi_1,\phi_2$ of the pairing potentials, related as $\phi_1-\phi_2=\phi$  and connected by a tunneling term, as indicated in the sketches of
Figs. \ref{fig3} and \ref{angb}.
Gauging out the dependence on $\phi$ in the operators of the wires, the Hamiltonian for the full system reads $H(\phi)=H_{\rm w1}+H_{\rm w2}+H_c(\phi)$, where $H_{\rm w1},\;H_{\rm w2}$ have
 the same structure as in Eq. (\ref{hamw}).
The connecting term reads
\begin{equation}\label{junct}
H_{\rm c}(\phi) = t_{\rm c} \sum_{\sigma=\uparrow, \downarrow} \left(e^{i \phi/2} c^{\dagger}_{1,\sigma} c_{2,\sigma} + \text{H.c.} \right),
\end{equation}
with $1$ and $2$ denoting, respectively, the site at the right/left end of ${\rm w1}/ {\rm w2}$. 
We can calculate the current numerically as described in the SM \cite{sm}.
In the topological phase, however, a simple description based
on the coupling of the MZMs accurately  explains the Andreev spectrum and 
 the CPR. This is because, in a topological junction, Andreev states 
are formed  from the hybridization of the MZMs  \cite{trs6,nos,cata1}.
\begin{figure}[h]
\begin{center}
\includegraphics[width=\columnwidth]{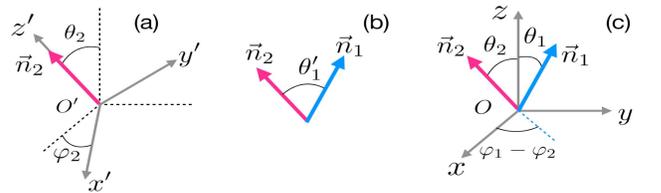}
\end{center}
\caption{(a) Reference frame with $\vec{z}^{\prime}$ along $\vec{n}_{2}$. 
(b) Bloch vectors of the MZMs of the two wires. (c) Laboratory frame.}
\label{sketch1}
\end{figure}

In what follows we derive the low-energy effective Hamiltonian $H_{\rm eff}$ that describes the hybridization of the MZMs. Importantly,
 we consider different magnetic-field and SOC orientations in the two wires.  $H_{\mathrm{eff}}$  takes a particularly
simple form if the quantization axis is chosen in the direction  of the 
Bloch vector of one of the MZMs next to the junction, which we choose to be $\vec{n}_{2}$. 
In the basis where $\vec{n}_{2}\equiv\vec{z}^{\prime}$ --- see Fig. \ref{sketch1} (a) ---
the spin down operators of the sites nearest to the junction
contribute only at high energies,  while the low-energy component is precisely the contribution of the MZM. Concretely, we can substitute the fermionic operators at the ends of the wires
by their projection on the MZMs, 
\begin{eqnarray}
c^{\prime}_{1\uparrow } &\simeq &\frac{a_{1}}{2} e^{i \delta^{\prime}_{1}}\cos
\left( \frac{\theta_{1}^{\prime}}{2}\right) \eta _{1},  \;\;\;\;\;\;\;\;\;\;
c^{\prime}_{2\uparrow } \simeq \frac{a_{2}}{2} e^{i \delta^{\prime}_{2}}\eta _{2},
\label{ceta}
\end{eqnarray}
where
 $\theta_1^{\prime}$ is the angle between $\vec{n}_1$ and $\vec{n}_2$ --- see Fig. \ref{sketch1} (b) --- and
 $\delta^{\prime}_1, \delta^{\prime}_2$ are the corresponding phases.
  $a_{i}$ are real numbers, $a_{i}^{2}\leq 1$ being the weight of the MZM at the corresponding site.
  Replacing in Eq. (\ref{junct}) we obtain
\begin{eqnarray}
H_{\mathrm{eff}}(\phi) &=&
\frac{t_{J}(\theta_1^{\prime})}{2} \sin \left( \frac{\phi }{2}+\delta^{\prime}_{2}-\delta^{\prime}_{1} \right) i \eta _{1}\eta _{2}, \label{ef1} \\
t_{J} (\theta_1^{\prime}) &=& t_{c}a_{1}a_{2}\cos \left( \frac{ \theta_1^{\prime}}{2}\right),\label{tj}
\label{ef2}
\end{eqnarray}
which is solved by defining a fermion 
$d=(\eta_{1}+i\eta _{2})/2$ \cite{kitaev}, leading to 
$i\eta _{1}\eta_{2}=2d^{\dagger } d-1$. The ground-state energy is 
\begin{eqnarray}
E_{\mathrm{eff}}(\phi ) &=&-\frac{1}{2}|t_{J}(\theta_1^{\prime})|
\left\vert \sin \left( \frac{\phi ^{\prime }}{2}\right) \right\vert,   \label{eg}
\end{eqnarray} 
where $\phi ^{\prime } =\phi +2\left( \delta^{\prime}_{2}-\delta^{\prime}_{1}\right)$. 
The CPR is
\begin{equation}
J_{\mathrm{eff}}(\phi ) =\frac{2e}{\hbar }
\frac{dE_{\mathrm{eff}}(\phi )}{d\phi }  
=-\frac{e|t_{J}(\theta_1^{\prime})|}{2\hbar }\cos \left( \frac{\phi ^{\prime }}{2}\right) 
\mbox{sgn} \left\{ \sin \left( \frac{\phi ^{\prime }}{2}\right) \right\}.  \label{jeff}
\end{equation}
Performing the rotation sketched in Fig. \ref{sketch1} (see SM \cite{sm} for details), we can express
this current in terms of the GBC of the MZMs of w1 and w2 next to the junction 
in the laboratory frame through
\begin{equation}
\phi ^{\prime }=\phi +2\left( \delta_{2}- \delta_{1} -\xi_{1,2} \right),
\label{phip}
\end{equation}
where  $\delta_1$ and $\delta_2$ are the corresponding phases and
\begin{equation}
\xi_{1,2} =\arctan \left[ \frac{\sin \left( \varphi _{1}-\varphi _{2}\right) }{\cos \left(
\varphi _{1}-\varphi _{2}\right) +\cot \left( \frac{\theta _{1}}{2}\right)
\cot \left( \frac{\theta _{2}}{2}\right) } \right]. 
\label{delta}
\end{equation}
The different angles are indicated in Fig. \ref{sketch1}. 
We would like to stress that all the quantities that determine the behavior of the CPR are 
 SU(2)-invariant, as explicitly shown in the SM \cite{sm}. In particular, $\theta _{1}^{\prime}$ does not depend on the reference frame while
  $\delta_{2} - \delta_{1} -\xi_{1,2}$ is an invariant akin to Eq. (\ref{rel}) and from Eq. (\ref{phip}) we clearly see that  this quantity plays the role of a vector potential that modifies the magnetic flux.

The CPR of Eq. (\ref{jeff}) has a jump at $\phi ^{\prime }=0$ as a consequence of the crossing of levels with different
fermion parity. 
If parity is conserved, the  typical $4\pi$-periodicity of 
topological junctions is obtained. 
In the case of junctions of wires with the same orientation of
$\vec{B}$ and SOC and $\vec{n}_B \cdot \vec{n}_{\lambda}=0$, $\delta_1-\delta_2=\pm \pi$, as given by Eq. (\ref{lr}), and the jump occurs at $\phi=\pi$. 
However, in junctions of wires having different orientations of
$\vec{n}_B$ and $\vec{n}_{\lambda}$ or $\vec{n}_B \cdot \vec{n}_{\lambda} \neq 0$, 
this jump may take place at other values of $\phi$. 
In what follows, we analyze  junctions of  wires with different configurations of these vectors
with the aim of using the behavior of the CPR to extract information of the MZMs.

We consider now the same orientation of $\vec{B}$ in both wires, but  a tilt $\beta_{\lambda}$ in the orientation of the SOC, i.e. $\vec{n}_{\lambda,1} \cdot \vec{n}_{\lambda,2} =\cos \beta_{\lambda}$. 
This can be realized with a junction where the wires are placed on the superconducting substrate 
forming an angle $\beta_{\lambda}$, as in the sketch of Fig. \ref{fig3}, where
we also indicate the ECF for w2 ($\vec{n}_B || \vec{z}$ and $\vec{n}_{\lambda,2}||\vec{x}$). 
We focus on $\Delta>0$, $\mu <0$, in which case  Eqs. (\ref{apl}) give $\delta_2$
and $\varphi_2$, while $\delta_1$ and $\varphi_1$ can be be also derived from these Eqs.
by performing a 
rotation of $\beta_{\lambda}$
around $\vec{z}$. This leads to
 $\theta_1=\theta_2=\theta$, $\varphi_1-\varphi_2=\pi
-\beta_{\lambda} $ and $\delta_{1}=\delta_{2}+(\pi +\beta_{\lambda} )/2$.
Replacing in Eqs.  (\ref{phip}) and    (\ref{delta}) we obtain
\begin{equation}
\phi ^{\prime }=\phi -\pi -\beta_{\lambda} +2\arctan \left[ \frac{\sin \left( \beta_{\lambda}
\right) }{\cos \left( \beta_{\lambda} \right) -\cot ^{2}\left( \frac{\theta }{2}
\right) }\right] .  \label{tso}
\end{equation}
Therefore, from the position of the jump in the current as a function of the flux it is possible to extract the angle $\theta$ between the Bloch vector of the MZMs with respect to $\vec{n}_B$. This completes the full description of the MZMs at both sides of the junction. In Fig. \ref{fig3}, we show results 
calculated with $H_{\rm eff}$,  and by  exact diagonalization of the full Hamiltonian $H(\phi)$ (see Ref \cite{sm} for technical details). Both calculations are in excellent  agreement and also agree with results reported in the limit of weak SOC in the continuum model \cite{klino2} and in the limit of large $B$ \cite{span}.

\begin{figure}[h]
\begin{center}
\includegraphics[width=\columnwidth]{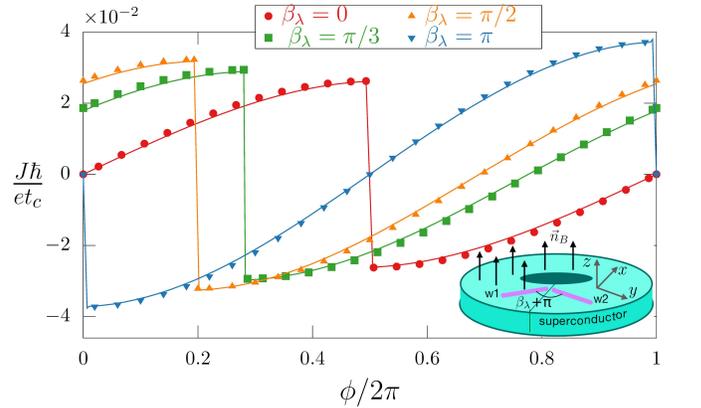}
\end{center}
\caption{CPR for $\vec{n}_{B,j} \cdot \vec{n}_{\lambda,j}=0,\; j=1,2$ and 
several values of the 
angle $\beta_{\lambda}$ between
$\vec{n}_{\lambda,1}$ and $\vec{n}_{\lambda,2}$.
Solid lines: numerical results. Symbols: $J_{\rm eff}$ calculated using $H_{\rm eff}$.
Parameters are  $t=1$, $B=4$, $\lambda=2$, $\Delta=2$ and $\mu=-3$.}
\label{fig3}
\end{figure}

We now focus on the case where the SOC is equally oriented  in the two wires, $\vec{n}_{\lambda,1}=\vec{n}_{\lambda,2}=\vec{x}$, while 
the orientation of the magnetic field $\vec{n}_{B,1}$ is tilted by an 
angle $ \beta_B$ with respect to $\vec{n}_{B,2} || \vec{z}$. We
start with the case $\vec{n}_{\lambda,j }\cdot \vec{n}_{B,j}=0, \; j=1,2$, which can be realized in the two configurations sketched in Fig. \ref{angb}. As before, for  $\Delta>0$, $\mu <0$,  Eqs. (\ref{apl}) give us the values of
$\delta _{2}$ and $\varphi _{2}$. On the other hand, the corresponding values of $\delta_1$ and $\varphi_1$ can also be obtained from these Eqs.
 by performing a 
rotation of angle $\beta_{B} $ around $\vec{x}$. These are
$\delta_{1}=-\pi /4$ and  
$\varphi _{1}=(\pi /2)\mbox{sgn}\lbrack \sin \left( \theta_2 -\beta_{B} \right) ]$ and $\theta_1=\theta_2-\beta_B$. 
Hence, the CPR is given by 
Eq. (\ref{jeff})  with $\phi ^{\prime }=\phi -\pi $. Therefore, the 
\emph{shape }of the function $J(\phi)$ is the same for all values of 
$\beta_B$, displaying a jump at 
$\phi=\pi $. However, the \emph{magnitude  } of the current depends on the angles $\theta_2$ and $\beta_B$ according to Eq. (\ref{tj}), with $\theta_1^{\prime}= 2 \theta_2-\beta_B $.
  This is illustrated in Fig. \ref{angb} and has a simple interpretation.
   For $\beta_B=0$, $\vec{n}_1$ and $\vec{n}_2$ have the same $z$ component, $\theta_1=\theta_2$,
zero $x$ component and opposite $y$ components [see Eqs. (\ref{lr}) and (\ref{apl})]. Rotating $\vec{n}_{B,1}$ around the $x$ axis, $\vec{n}_1$ is moved towards $\vec{n}_2$ and both vectors coincide when $\beta_B=2 \theta_2$. This angle corresponds to the maximum of $t_J(\theta_1^{\prime})$, hence, the maximum of  $J(\phi)$ at fixed $\phi$. 
In addition, for fixed fermion parity, the CPR is $4\pi$-periodic in
$\beta_B$, in agreement with Ref. \onlinecite{trs6m}.

\begin{figure}[h]
\begin{center}
\includegraphics[width=\columnwidth]{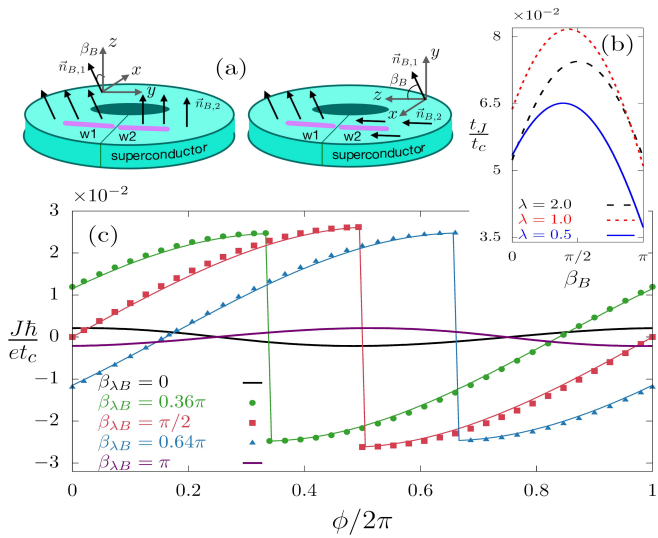}
\end{center}
\caption{(a) Configurations of wires with a tilt 
$\beta_B$ in the orientation of the magnetic field 
with $\vec{n}_B \cdot \vec{n}_{\lambda}=0$. 
(b)
Amplitude of the CPR $t_J$ vs. 
$\beta_B$ for $\vec{n}_B \cdot \vec{n}_{\lambda}=0$. 
(c) CPR for 
$\vec{n}_{\lambda,1}=\vec{n}_{\lambda,2}$,
$\vec{n}_{B,2}\cdot \vec{n}_{\lambda,2}=0$, 
$\vec{n}_{B,1}\cdot \vec{n}_{\lambda,1}=\cos{\beta_{\lambda B}}$ and all vectors in the same plane. 
Parameters as in Fig. \ref{fig3}.}
\label{angb}
\end{figure}

When $\vec{B}_1$ is  tilted in such a way that there is a finite component along the direction of the SOC there is no simple analytical expression relating the tilt in $\vec{B}_1$ and the orientation of
the Bloch vector of the MZMs and we must rely on the full expressions given by Eqs. (\ref{jeff}), (\ref{phip}) and (\ref{delta}). 
In Fig. \ref{angb} (c) we show the CPR for the case in which
$\vec{n}_{\lambda,1}=\vec{n}_{\lambda,2}$,
$\vec{n}_{B,2}\cdot \vec{n}_{\lambda,2}=0$ and  $\vec{n}_{B,1}$ is tilted keeping it perpendicular to 
$\vec{n}_{B,2} \wedge \vec{n}_{\lambda,2}$, and forming an angle $\beta_{\lambda B}$ 
with $\vec{n}_{\lambda,1}$.
Without tilting, $J(\phi)=0$ as expected \cite{note3}. 
For other cases, $J(\phi)$
presents jumps at $\phi\neq \pi$ as in the case of wires with  SOC perpendicular to $B$ 
but with a relative tilt, analyzed in Fig. \ref{fig3}.
 It is  found again an excellent agreement between the description in terms of the effective
Hamiltonian $H_{\rm eff}(\phi)$ and the numerical solution of the exact Hamiltonian (see SM \cite{sm} for details). 
For small $\beta_{\lambda B}$, the topological phase is lost in ${\rm w1}$ 
leaving its place to a non-topological phase --- see of Eq. (\ref{bound}) --- which is gapless for a wide  parameter range. There,
$H_{\rm eff}(\phi)$ is no longer useful and the numerical solution of $H(\phi)$ is necessary, which leads to a CPR, typical of ordinary superconductors, with small amplitude, albeit preserving some peculiar features of the topological phase, like $J(0)\neq 0$ \cite{note3}, similar to Ref. \cite{eggernt}. 

We have characterized the MZMs of topological superconducting wires with SOC and 
magnetic field in terms of GBC  $(\varphi, \theta,\delta)$. 
We have analytically calculated $\varphi, \delta$ for the ECF where $\vec{n}_B \equiv \vec{z}$ and $\vec{n}_{\lambda} \equiv \vec{x}$. We have also derived  the transformation of these quantities under changes of the reference frame. We used these results to derive exact expressions for the CPR in wires having relative tilts in the  orientations of the SOC and magnetic fields.  We showed that for suitable configurations of the junctions, the CPR provides the necessary information to fully reconstruct the structure of the MZMs. These results may be useful in the experimental implementation of quantum tomography of MZMs.  The dc regime could be reached, for instance, by adiabatically switching on the magnetic field or by rotating it from the
gapless non-topological phase of 
nearly parallel SOC and magnetic field. This 
 is possible within the present experimental state of the art of the hybrid superconducting-semiconducting wires we have studied \cite{wires-jose}. Interestingly, this regime is free from 
  the problem of the time-scales  introduced by the poor equilibration of the MZMs which affect readout processes of dynamical effects \cite{roy,seoane,tuovin}.

LA thanks A. Levy Yeyati  and F. von Oppen for stimulating discussions. We acknowledge support from CONICET, Argentina  and  the Alexander von Humboldt
Foundation, Germany (LA). We are sponsored by  PIP-RD 20141216-4905 and PIP 112-201501-00506 
of CONICET, 
 PICT-2017-2726, PICT-2018-04536 and PICT-Raices-2018.


\newpage
\begin{widetext}

\setcounter{equation}{0}
\setcounter{figure}{0}
\setcounter{table}{0}
\setcounter{page}{1}

\makeatletter
\renewcommand{\theequation}{S\arabic{equation}}
\renewcommand{\thefigure}{S\arabic{figure}}
\renewcommand{\bibnumfmt}[1]{[S#1]}
\renewcommand{\citenumfont}[1]{S#1}
\begin{center}
\large{\bf Supplemental Material: Tomography of zero-energy end modes in topological superconducting wires}
\end{center}

\maketitle


\maketitle

\setcounter{equation}{0} \setcounter{figure}{0} \setcounter{table}{0} 
\setcounter{page}{1}

\makeatletter
\renewcommand{\theequation}{S\arabic{equation}} \renewcommand{\thefigure}{S\arabic{figure}} \renewcommand{\bibnumfmt}[1]{[S#1]} \renewcommand{\citenumfont}[1]{S#1}


\section{Change of reference frame}
\subsection{ General case }\label{change}

The spin of the fermionic creation operators defined in Eq. (4) of the main text is expressed in a given reference frame $O$, determined by
the quantization axis of the Hamiltonian $H_{\rm w}$. Here, we analyze the
transformation of the spin under a change of basis to a rotated frame 
$O^{\prime }$. 
We remind the reader that under an active transformation, (a rotation of the physical system an angle $\alpha $ around de unit vector $\vec{v}$  keeping the coordinates unchanged) a state 
$|\psi \rangle  = (a c_{\uparrow }^{\dagger}
+ b c_{\downarrow }^{\dagger })|0\rangle )$, becomes 
$R_{\vec{v}}(\alpha ) |\psi \rangle $, 
where the SU(2) matrix $R_{\vec{v}}(\alpha )$ is
\begin{equation}
	R_{\vec{v}}(\alpha )=\cos \left(\frac{\alpha }{2}\right)-i\sin\left( \frac{\alpha }{2}\right)
\vec{v}\cdot \vec{\sigma}.  \label{su2}
\end{equation}
If, instead, the physical system is fixed and the rotation is applied to the
coordinate system $O$ to transform it to $O^{\prime }$, the 
state in the
new basis is $|\psi ^{\prime }\rangle =R_{\vec{v}}^{-1}(\alpha )|\psi \rangle$.
Inverting the previous transformation we obtain for the creation operators 
\begin{eqnarray}
	c_{\uparrow }^{\dagger } &=&\left( \cos \left(\frac{\alpha }{2} \right) +e^{-i\frac{\alpha}
	{2}}v_{z}\right) ({c}_{\uparrow }^{\prime })^{\dagger }+\sin \left( \frac{\alpha }{2}\right)
\left( v_{y}-iv_{x}\right) ({c}_{\downarrow }^{\prime })^{\dagger }  \notag
\\
	c_{\downarrow }^{\dagger } &=&-\sin \left(\frac{\alpha }{2}\right)\left(
v_{y}+iv_{x}\right) ({c}_{\uparrow }^{\prime })^{\dagger }
	+\left( \cos\left( \frac{\alpha }{2}\right)+v_{z}e^{i\frac{\alpha}{ 2}}\right) 
({c}_{\downarrow }^{\prime})^{\dagger },  \label{tra}
\end{eqnarray}
where $v_{j}$ is the component of $\vec{v}$ in the direction $j$. Replacing
this transformation in Eq. (4) of the main text we obtain the expression of the creation component
of the MZM in the rotated frame $O^{\prime }$: 
\begin{eqnarray}
\gamma _{\nu }^{\dagger } &=&e^{i\delta _{\nu }}\left[ A({c}_{\uparrow
}^{\prime })^{{\dagger }}+B({c}_{\downarrow }^{\prime })^{\dagger }\right] ,
\notag \\
	A &=&\cos\left (\frac{\theta _{\nu }}{2}\right)\left( \cos\left( \frac{\alpha }{2}\right)+e^{-i\frac{\alpha}
	{2}}v_{z}\right) -e^{i\varphi _{\nu }}\sin \left(\frac{\theta _{\nu }}{2}\right)\sin \left(\frac{\alpha}
	{2}\right)\left( v_{y}+iv_{x}\right) ,  \notag \\
	B &=&\cos \left(\frac{\theta _{\nu }}{2}\right)\sin \left(\frac{\alpha}{2}\right)\left( v_{y}-iv_{x}\right)
	+e^{i\varphi _{\nu }}\sin \left(\frac{\theta _{\nu }}{2}\right)\left( \cos \left(\frac{\alpha }{2}\right)
	+v_{z}e^{i\frac{\alpha}{2}}\right) .  \label{ab}
\end{eqnarray}

Expressing $\gamma _{\nu }^{\dagger }$ in the same form as Eq. (4) of the main text we get
\begin{equation}
	\gamma _{\nu }^{\dagger }=e^{i\delta _{\nu }^{\prime }}\left[ \cos \left(\frac{\theta
	_{\nu }^{\prime }}{2}\right)({c}_{\nu \uparrow }^{\prime })^{\dagger }+e^{i\varphi
	_{\nu }^{\prime }}\sin \left(\frac{\theta _{\nu }^{\prime }}{2}\right)({c}_{\nu \downarrow
}^{\prime })^{\dagger }\right] ,  \label{cre}
\end{equation}
from where the parameters in the frame $O^{\prime }$ can be obtained. Writing 
$A=|A|e^{i\xi_{\nu } }$, it is clear that  $\delta _{\nu }^{\prime }=\xi_{\nu }
+\delta _{\nu }$ being 
\begin{equation}
	\xi_{\nu } =\arctan\left( \frac{\mbox{Im}A}{\mbox{Re}A}\right)
	=\arctan\left( \frac{-\cos (\frac{\theta _{\nu }}{2})\sin (\frac{\alpha}{
	2})v_{z}-\sin (\varphi _{\nu })\sin (\frac{\theta _{\nu }}{2})\sin (\frac{\alpha}{
		2})v_{y}-\cos (\varphi _{\nu })\sin (\frac{\theta _{\nu }}{2})\sin (\frac{\alpha}{2})v_{x}}
	{\cos (\frac{\theta _{\nu }}{2})\cos (\frac{\alpha}{2})(1+v_{z})-\cos (\varphi _{\nu })\sin
	(\frac{\theta _{\nu }}{2})\sin (\frac{\alpha}{2})v_{y}+\sin (\varphi _{\nu })\sin (\frac{\theta
	_{\nu }}{2})\sin (\frac{\alpha}{2})v_{x}}\right) .  \label{delta}
\end{equation}
We see that in general, the phases $\delta _{\nu }$ transform in a non
trivial way under rotations or a change in coordinates. Instead, as
expected, the directions $\vec{n}_{\nu }$ (defined by $\theta _{\nu }$ and 
$\varphi _{\nu }$) transform as ordinary vectors. Comparing Eqs. (\ref{ab})
and (\ref{cre}) we see that $B/A=|B/A|e^{i\varphi _{\nu }^{\prime }}$ or 
$\bar{A}B=|\bar{A}B|e^{i\varphi _{\nu }^{\prime }}$ ($\bar{A}$ denotes the complex conjugate of $A$), and $|A|=\cos (\theta
_{\nu }^{\prime }/2)$, from which $\theta _{\nu }$ and $\varphi _{\nu }$ are
easily obtained:
\begin{equation}
\theta _{\nu }^{\prime }=2\arccos (|A|)=2\arctan (|B/A|).  \label{theta}
\end{equation}
\begin{equation}
\varphi _{\nu }^{\prime }=\arctan 
\left( \frac{\mbox{Im}(\bar{A}B)}{\mbox{Re}(\bar{A}B)}\right) ,  \label{fi}
\end{equation}

\subsection{Derivation of Eq. (18)}
\label{chjose}
In the main text, we evaluate the Josephson  current through the connection between wires ${\rm w}1$ and ${\rm w}2$, with the parameters defined with respect to a
frame $O^{\prime}$ with $\vec{n}_{2}||\vec{z}^{\prime}$, being $\vec{n}_2$ the direction of the polarization of the MZM of the wire ${\rm w 2}$ that hybridizes with the MZM of the wire
${\rm w1}$  in the junction. The consequent expression for the Josephson current --see Eq. (16) of the main text-- depends on the Josephson phase $\phi$, as well as on the  phases $\delta^{\prime}_1$ and 
$\delta^{\prime}_2$ of the two hybridized MZMs, which depend on the reference frame. Since we know the values of these phases, given the values of the parameters of the Hamiltonians for the wires only when the latter are written in
the reference frame $O$ where $\vec{n}_B || \vec{z}$ and $\vec{n}_{\lambda} ||\vec{x}$ --see Eq. (10)-- we need to implement a transformation between $O^{\prime}$ and $O$. The concrete transformation is sketched 
 in Fig. 1 of the main text.
In the formalism described above, this corresponds to a 
rotation $R_{\vec{v}}(\alpha )$ that transforms $O$ to $O^{\prime }$ such that
 $R_{\vec{v}}(\alpha )\vec{n}_{2}=\vec{z}$.  We choose  
$\vec{v}$ in the direction of  $\vec{n}_{2}\wedge \vec{z}$, so that it is
perpendicular to both $\vec{n}_{2}$ and $\vec{z}$, hence  a positive
rotation in the angle $\alpha =\theta _{2}$ moves $\vec{n}_{2}$ to $\vec{z}$. 
The components of the unit vector $\vec{v}$ become 
$v_{x}=\sin (\varphi_{2})$, $v_{y}=-\cos (\varphi _{2})$, $v_{z}=0.$

Replacing these values in Eq. (\ref{delta}) for $\nu=2$, we see that the
numerator vanishes, and therefore $\xi_2=0$, $\delta _{2}^{\prime }=\delta _{2}$. 
Instead for $\nu =1$ we obtain $\delta _{1}^{\prime }=\xi_1 +\delta _{1}$,
with 
\begin{equation}
\xi_1 =\arctan\left( \frac{\sin \left( \varphi _{1}-\varphi
_{2}\right) }{\cot \left( \frac{\theta _{1}}{2}\right) \cot 
\left( \frac{\theta _{2}}{2}\right) +\cos \left( \varphi _{1}
-\varphi _{2}\right) }\right).  \label{deltade1}
\end{equation}
Combining the $\delta_2^{\prime}- \delta_1^{\prime}$, we get Eqs. (17) and (18) of the main text, with $\xi_{1,2}\equiv \xi_1$, given above.

\subsection{SU(2) invariance of $d_{12}$}
\label{inv}

In this section we prove the SU(2) invariance of the quantity
\begin{equation}
d_{12}=\delta _{1}-\delta _{2}-\xi_{1,2}, \;\;\;\;\;\;\;\; \xi_{1,2}=
\arctan \left( \frac{\sin \left( \varphi
_{1}-\varphi _{2}\right) }{\cot \left( \frac{\theta _{1}}{2}\right) \cot
\left( \frac{\theta _{2}}{2}\right) +\cos \left( \varphi _{1}-\varphi
_{2}\right) }\right) ,  \label{d12}
\end{equation}
mod$(\pi )$ for any two fermions of  the form of Eq. (4) of the main text
[same as Eq. (\ref{cre}) without the superscript prime]. The fact that the quantity
is defined mod$(\pi )$ means that the branch and discontinuities of the $\arctan $ 
are unimportant. 
The   invariance of $d_{12}$ is expected, since  
in the particular case discussed in Section \ref{chjose},
it enters the equation
of the Josephson current through $\phi ^{\prime }$ [see Eqs. (17) of the
main text] and 
the current is an observable. Here, we prove it explicitly for the general case. 

As is well known, any SU(2) rotation can be obtained by composing
infinitesimal rotations around three mutually perpendicular axis and the
generators of these rotations ($i\sigma _{x}$, $i\sigma _{y}$ and $i\sigma
_{z}$ in Section \ref{change})  form a basis of the Lie algebra of the group.
Two generators are enough for our purposes because the third one is the
commutator of the other two times a factor. The invariance of $d_{12}$ under
any rotation around $z$ immediately verified since $\theta_1$ and $\theta_2$, as well as  $\delta _{1}-\delta _{2}$ and 
$\varphi _{1}-\varphi _{2}$ are unchanged under this transformation.
Therefore, it remains to prove that $d_{12}$ is invariant under a rotation
through an axis perpendicular to $z$. We choose the $y$ axis in a reference frame with 
$\varphi _{2}=0$ to simplify the calculation (the axis forming an
angle $\pi /2+\varphi _{2}$ with the $x$ axis in the original reference frame).

We use the results of Section \ref{change} for $\vec{v}=\vec{y}$, 
$\varphi_{2}=0$ and $\alpha \rightarrow 0$ to linear order in the differential 
$d\alpha $ of the angle of the rotation. In particular\ we replace 
$\cos(\alpha /2)\simeq 1$ and $\sin (\alpha /2)\simeq d\alpha /2$. From Eq. 
(\ref{delta}) we obtain the change in the phase under the infinitesimal rotation, $d\delta _{\nu}=\delta _{\nu }^{\prime }-\delta _{\nu }$,
\begin{equation}
d\delta _{\nu}=d\arctan (\delta _{\nu})=-\frac{\frac{d\alpha }{2}\sin (\varphi_{\nu})\sin (\frac{\theta _{\nu}}{2})}{\cos (\frac{\theta _{\nu}}{2})}.  \label{di}
\end{equation}
Evaluating explicitly for $\nu=1,2$ this equation reads
\begin{equation}
	\frac{d\delta _{1}}{d\alpha }=-\frac{1}{2}\tan \left(\frac{\theta _{1}}{2}\right)\sin (\varphi
_{1}),\text{ }\;\;\;\;\;\;\;\; \frac{d\delta _{2}}{d\alpha }=0.  \label{di2}
\end{equation}
From Eqs. (\ref{ab}) and (\ref{theta}) we get
\begin{equation}
\cot \left( \frac{\theta _{\nu}^{\prime }}{2}\right) ^{2}=\cot 
\left( \frac{\theta _{\nu}}{2}\right) ^{2}\frac{1-\tan (\frac{\theta _{\nu}}{2})\cos (\varphi_{\nu})d\alpha }{1+\cot (\frac{\theta _{\nu}}{2})\cos (\varphi _{\nu})d\alpha }
=\cot \left( \frac{\theta _{\nu}}{2}\right) ^{2}\left( 1-\cos (\varphi_{\nu})d\alpha \left( \tan \left( \frac{\theta _{\nu}}{2}\right) +\cot \left( \frac{\theta _{\nu}}{2}\right) \right) \right)  \label{cot1}
\end{equation}
\begin{equation}
d\cot \left( \frac{\theta _{\nu}}{2}\right) ^{2}=-\cot 
\left( \frac{\theta _{\nu}}{2}\right) ^{2}\left( \tan \left( \frac{\theta _{\nu}}{2}\right) +\cot \left( \frac{\theta _{\nu}}{2}\right) \right) \cos (\varphi _{\nu})d\alpha. 
\end{equation}
Using  that for any function $r$, $dr^{2}=2rdr$ we obtain
\begin{equation}
\frac{d\cot (\theta _{1}/2)}{d\alpha }=-\frac{\cos (\varphi _{1})}{2}\left(
1+\cot \left( \frac{\theta _{1}}{2}\right) ^{2}\right)  \label{eq:dc1}
\end{equation}
\begin{equation}
\frac{d\cot (\theta _{2}/2)}{d\alpha }=-\frac{1}{2}\left( 1+\cot \left( 
\frac{\theta _{2}}{2}\right) ^{2}\right)  \label{eq:dc2}
\end{equation}
The change of the angles $d\varphi _{\nu}=\varphi _{\nu}^{\prime }-\varphi _{\nu}$
are obtained using Eqs. (\ref{ab}) and (\ref{fi})
\begin{equation}
\tan (\varphi _{\nu}^{\prime })=\frac{\sin (\varphi _{\nu})\sin (\frac{\theta_{\nu}}{2})\cos (\frac{\theta _{\nu}}{2})}{\cos (\varphi _{\nu})\sin (\frac{\theta_{\nu}}{2})\cos (\frac{\theta _{\nu}}{2})+(\cos (\frac{\theta _{\nu}}{2})^{2}-\sin
(\frac{\theta _{\nu}}{2})^{2})\frac{d\alpha }{2}}
\end{equation}
\begin{equation}
d\tan (\varphi _{i})=-\frac{d\alpha \sin (\varphi _{i})}{2\cos (\varphi
_{i})^{2}}\left( \cot \left( \frac{\theta _{i}}{2}\right)
-\tan \left( \frac{\theta _{i}}{2}\right) \right) .
\end{equation}
Using $d\tan (r)=(1+\tan (r)^{2})dr$ 
\begin{equation}
\frac{d\varphi _{1}}{d\alpha }=-\sin (\varphi _{1})\left( \cot 
\left( \frac{\theta _{1}}{2}\right) -\tan \left( \frac{\theta _{1}}{2}\right) \right) ,
\text{ }\frac{d\varphi _{2}}{d\alpha }=0.  \label{eq:dphi1}
\end{equation}

The remaining task to prove that $dd_{12}/d\alpha =0$ is to derive $\xi_{1,2} =\arctan (q)$, where 
\begin{equation}
q=\frac{\sin (\varphi _{1})}{\cos (\varphi
_{1})+\cot (\frac{\theta _{1}}{2})\cot (\frac{\theta _{1}}{2})}  \label{eq:q}
\end{equation}
To simplify the algebra we use the notation $c=\cos \left( \varphi
_{1}\right) $, $s=\sin \left( \varphi _{1}\right) $ and $x_{i}=\cot \left(
\theta _{i}/2\right) $ . With this notation the equations (\ref{eq:dc1}), 
(\ref{eq:dc2}), (\ref{eq:dphi1}) and (\ref{eq:q}) become
\begin{equation}
\frac{dx_{1}}{d\alpha }=-\frac{c}{2}\left( 1+x_{1}^{2}\right) ,
\text{ }\frac{dx_{2}}{d\alpha }=-\frac{1}{2}\left( 1+x_{2}^{2}\right) ,
\text{ }\frac{d\varphi _{1}}{d\alpha }=-\frac{s}{2}\left( x_{1}-\frac{1}{x_{1}}\right),\text{ }q=\frac{s}{c+x_{1}x_{2}}=\frac{s}{h}.
\label{eq:dx1}
\end{equation}
Differentiating the last expression we get
\begin{equation}
\frac{dq}{d\alpha }=\frac{c\frac{d\varphi _{1}}{d\alpha }}{h}-\frac{s\left(
-s\frac{d\varphi _{1}}{d\alpha }+\frac{dx_{1}}{d\alpha }x_{2}
+x_{1}\frac{dx_{2}}{d\alpha }\right) }{h^{2}}=\frac{\frac{d\varphi _{1}}{d\alpha }
+cx_{1}x_{2}\frac{d\varphi _{1}}{d\alpha }
-s\left( \frac{dx_{1}}{d\alpha }x_{2}+x_{1}\frac{dx_{2}}{d\alpha }\right) }{h^{2}}
\end{equation}
and replacing Eqs. (\ref{eq:dx1}) above, we obtain
\begin{equation}
\frac{dq}{d\alpha }=\frac{s}{2x_{1}}\frac{1+2cx_{1}x_{2}+x_{1}^{2}x_{2}^{2}}{h^{2}}  \label{qpri}
\end{equation}
On the other hand, from Eq. (\ref{eq:q})
\begin{equation}
\frac{d\xi_{1,2} }{d\alpha }=\frac{\frac{dq}{d\alpha }}{1+q^{2}},\text{ with } 1+q^{2}=1+\frac{s^{2}}{h^{2}}=\frac{1+2cx_{1}x_{2}+x_{1}^{2}x_{2}^{2}}{h^{2}},
\end{equation}
and using Eq. (\ref{qpri}) we obtain
\begin{equation}
\frac{d\xi_{1,2} }{d\alpha }=\frac{s}{2x_{1}}=\frac{\sin (\varphi _{1})}{2\cot
\left( \theta _{1}/2\right) }.  \label{dxi}
\end{equation}
Finally, differentiating Eq. (\ref{d12}) and expressing it as 
\begin{equation}
\frac{d d_{12}}{d\alpha} = \frac{d \delta_1}{d\alpha} - \frac{d \delta_2}{d\alpha}+\frac{d \xi_{1,2}}{d\alpha},
\end{equation}
and substituting Eqs. (\ref{di2}), (\ref{eq:q}), and (\ref{dxi})
we get the desired result
\begin{equation}
\frac{dd_{12}}{d\alpha }=0.
\end{equation}

\section{Structure of the Majorana states in some limiting cases}
\subsection{Solution for dominant spin-orbit coupling with  $\vec{n}_B \equiv \vec{x}$ and $\vec{n}_{\lambda} \equiv \vec{z}$}\label{cont}
We apply the formalism of Section \ref{change} to the exact solution of the continuum version of the model of Eq. (1) of the main text, calculated in Ref. [\onlinecite{wires1}]. A very simple expression was found for the left and right MZMs in the region of parameters where  the spin-orbit coupling dominates,  assuming $\Delta >0$, $\lambda \gg t$, $B>\Delta$, $\mu \sim 0$ (equivalent to $\mu \sim -2t$ in the lattice version). From there, we can easily examine the properties summarized in Eqs. (6) to (8) of the main text. 
The solution, as expressed in Ref. [\onlinecite{wires1}] reads
\begin{eqnarray}
\eta_{L}&=& \frac{1}{2} \left( \psi_{L,\uparrow} - i \psi_{L,\downarrow} + i \psi^{\dagger}_{L,\downarrow} + \psi^{\dagger}_{L,\uparrow} \right), \;\;\;\;
\eta_R =  \frac{1}{2} \left(\psi_{R,\uparrow} + i \psi_{R,\downarrow} -i \psi^{\dagger}_{R,\downarrow} + \psi^{\dagger}_{R,\uparrow} \right),\label{etas}
\end{eqnarray}
where the labels $L,R$ in the field operators indicate that they are evaluated at spacial coordinates corresponding the the $L, R$ ends, respectively.
 In order to make an explicit comparison to Eqs. (7) and (8), we need to perform a rotation of $=\pi/2$ around the $y$- axis, corresponding to 
 $\alpha=-\pi/2$ and $\vec{v}=(0,1,0)$ in Eq. (\ref{tra}), and a change 
 in the sign of $\lambda$ which changes the sign of both $\delta$ and $\phi$ (see Ref. 53 of the main text).
 Under these transformations, the above operators transform to
 \begin{eqnarray}
 \gamma_{L}^{\prime}&=&  e^{i \pi/4} \left( \psi^{\prime}_{L,\uparrow} - i \psi^{\prime}_{L,\downarrow} \right), \;\;\;\;\;
\gamma_R^{\prime} =  e^{-i \pi/4} \left(\psi^{\prime}_{R,\uparrow} + i \psi^{\prime}_{R,\downarrow} \right),
 \end{eqnarray}
 in full agreement with Eqs. (7) and (8) of the main text.

\subsection{Solution for dominant magnetic field, $B\gg \Delta \gg \lambda$ with $\vec{n}_B \equiv \vec{z}$ and $\vec{n}_{\lambda} \equiv \vec{x}$}
\label{ala}

In this Section,  we obtain analytically the zero-energy modes at the ends of a
finite long chain for $0<\lambda \ll \Delta \ll t<B$ and $\mu \sim -B$. We
start with the Hamiltonian Eq. (9) of the main text, which to linear order
in $\lambda /B$ takes the form
\begin{equation}
H=\sum_{k,s =+,-}(-2t\cos k-\mu )\;d_{ks}^{\dagger }d_{ks
}-B\sum_{k}(d_{k+}^{\dagger }d_{k+}-d_{k-}^{\dagger }d_{k-})+\sum_{k}\left(
\Delta _{S}\;d_{k+}^{\dagger }d_{-k-}^{\dagger }-\Delta _{T}\sin k\sum_{s
=+,-}d_{ks}^{\dagger }d_{-ks }^{\dagger }+\text{H.c.}\right) ,
\label{hk}
\end{equation}
with $\Delta _{S}=\Delta $ and $\Delta _{T}=\lambda \Delta /B$. Transforming Fourier to Wannier functions  localized at any site $j$, 
$d_{j s }^{\dagger }=\sum_{k}e^{ijk}d_{ks }^{\dagger }/\sqrt{N}$, the
Hamiltonian becomes
\begin{equation}
H=-t\sum_{j,s =+,-}(d_{js }^{\dagger }d_{j+1s }+\text{H.c.})
-B\sum_{j}(d_{j+}^{\dagger }d_{j+}-d_{j-}^{\dagger }d_{j-})+\sum_{j}\left(
\Delta _{S}\;d_{j+}^{\dagger }d_{j-}^{\dagger }
+i\Delta _{T}\sum_{s
=+,-}d_{j+1s }^{\dagger }d_{js }^{\dagger }+\text{H.c.}\right) .
\label{hr}
\end{equation}
For later use we note that in the real-space basis, to linear order in $\lambda /B$ the transformation introduced in the main text to define Eq. (9) from Eq. (1) reads
\begin{equation}
d_{j+}^{\dagger }=c_{j\uparrow }^{\dagger }-\frac{i\lambda }{2B}\left(
c_{j+1\downarrow }^{\dagger }-c_{j-1\downarrow }^{\dagger }\right) ,\;\;\;\;\text{ }
d_{j-}^{\dagger }=c_{j\downarrow }^{\dagger }+\frac{i\lambda }{2B}\left(
c_{j+1\uparrow }^{\dagger }-c_{j-1\uparrow }^{\dagger }\right) .  \label{dac}
\end{equation}
In order to eliminate the imaginary unit in the coefficient $i\Delta _{T}$ of
the triplet superconductivity in Eq. (\ref{hr}) we define
\begin{equation}
\tilde{d}_{j+}^{\dagger }=e^{i\pi /4}d_{j+}^{\dagger },
\text{ }\tilde{d}_{j-}^{\dagger }=e^{-i\pi /4}d_{j-}^{\dagger }  \label{dtilde}
\end{equation}
and the triplet superconducting term takes the form
$\Delta _{T}\sum_{j}(\tilde{d}_{j+1 + }^{\dagger }\tilde{d}_{j + }^{\dagger }
-\tilde{d}_{j+1 - }^{\dagger }\tilde{d}_{j - }^{\dagger }
+\text{H.c.}) .
$

We obtain the solutions with zero energy of Eq. (\ref{hr}) for a finite long
chain of $N$ sites using the method of  
Alase \textit{et al.} \cite{alas1,alas2} in the form used previously 
by some of us.\cite{entangle}  As
in the Nambu formalism, the operators are mapped to one particle states,
using the following notation
\begin{equation}
\tilde{d}_{j s }\leftrightarrow |j s 1\rangle ,\text{ }\tilde{d}_{j s
}^{\dagger }\leftrightarrow |j s 2\rangle .  \label{map}
\end{equation}
The desired solutions are  linear combinations of states of the form (not
normalized)
\begin{equation}
|z s i\rangle =\sum_{j=1}^{N}z^{j-1}|j s i\rangle ,  \;\;\;\;\;\; s=\pm, \; i=1,2, \label{bloch}
\end{equation}
where  $z$ is a complex number with $|z|<1$ ($>1$) for the Majorana zero mode localized at the left (right) of
the chain. Since both modes are related by symmetry we focus here on the
left mode only. The possible values of $z$ are obtained from the bulk equation 
$P_{B}(H-E)|\psi \rangle =0$, where in our case $E=0$ and  
$P_{B}=\sum_{j=2}^{N-1}\sum_{s i}|j s i\rangle \langle j s i|$. In the
basis $|z,+,1\rangle $, $|z,+,2\rangle $, $|z,-,1\rangle $, $|z,-,2\rangle $,  
the matrix $P_{B}H$ takes the form
\begin{equation}
P_{B}H=\left( 
\begin{array}{cccc}
-a & -b & 0 & \Delta _{S} \\ 
b & a & -\Delta _{S} & 0 \\ 
0 & -\Delta _{S} & -a+2B & b \\ 
\Delta _{S} & 0 & -b & a-2B
\end{array}
\right) ,\text{ }a=\mu +B+t\left( z+\frac{1}{z}\right) ,
\text{ }b=\Delta_{T}\left( z-\frac{1}{z}\right)   \label{bulk}
\end{equation}
and its determinant is
\begin{equation}
\text{Det}(P_{B}H)=(a^{2}-b^{2})\left[ (a-2B)^{2}-b^{2}\right] -\left[
2a(2B-a)+2b^{2}\right] \Delta _{S}^{2}+\Delta _{S}^{4}.  \label{det}
\end{equation}
To linear order in $\Delta _{S}/B,$ we can neglect $\Delta _{S}$ above and
the four roots $z_{k}$ of Det$(P_{B}H)=0$ with $|z_{k}|<1$ and the
corresponding coefficients of the eigenvectors $|e_{k}\rangle =\sum_{s
i}\beta _{s i}^{k}|j s i\rangle ,$ for $\mu ^{\prime }=\mu +B\ll t$ are
\begin{eqnarray}
z_{1} &=&ic-\frac{\mu ^{\prime }}{2(t+\Delta _{T})},\text{ }\beta
_{+1}^{1}=\beta _{+2}^{1}=\frac{1}{\sqrt{2}},\text{ }\beta _{-1}^{1}=\beta
_{-2}^{1}=0,\text{ }c=\sqrt{\frac{t-\Delta _{T}}{t+\Delta _{T}},}  \notag \\
z_{2} &=&\bar{z}_{1}=-ic-\frac{\mu ^{\prime }}{2(t+\Delta _{T})},\text{ }
\beta _{s i}^{2}=\beta _{s i}^{1},  \notag \\
z_{3} &=&\frac{2B-\mu ^{\prime }}{2(t+\Delta _{T})}
-\sqrt{\left( \frac{2B-\mu ^{\prime }}{2(t+\Delta _{T})}\right) ^{2}-\frac{t-\Delta _{T}}{t+\Delta _{T}}},\text{ }\beta _{+1}^{3}=\beta _{+2}^{3}=0,\text{ }\beta
_{-1}^{3}=\beta _{-2}^{3}=\frac{1}{\sqrt{2}},  \notag \\
z_{4} &=&\frac{2B-\mu ^{\prime }}{2(t-\Delta _{T})}
-\sqrt{\left( \frac{2B-\mu ^{\prime }}{2(t-\Delta _{T})}\right) ^{2}-\frac{t+\Delta _{T}}{t-\Delta _{T}}},
\text{ }\beta _{+1}^{3}=\beta _{+2}^{3}=0,\text{ -}\beta
_{-1}^{3}=\beta _{-2}^{3}=\frac{1}{\sqrt{2}}.  \label{zk}
\end{eqnarray}
The zero mode state has the form $|f\rangle =\sum_{k}\alpha
_{k}|e_{k}\rangle $, and the coefficients are obtained from the boundary
equation, which in our case takes the form $P_{1}H|f\rangle =0$, where 
$P_{1}=\sum_{s i}|1s i\rangle \langle 1s i|$. It is easy to see that
the form of the matrix $P_{1}H$ is similar to Eq. (\ref{bulk}) without the
terms in $1/z$ (due to the fact that there are no sites at the left of site
1), and $z$ replaced by $z_{k}$. Taking for the basis state $|b\rangle $,
the four states $|z,+,1\rangle $, $|z,+,2\rangle $, $|z,-,1\rangle $, 
$|z,-,2\rangle $, $\langle b|P_{1}H|f\rangle =0$ imply
\begin{eqnarray}
\sum_{k}\left[ -\left( \mu ^{\prime }+tz_{k}\right) \beta _{+1}^{k}-\Delta
_{T}z_{k}\beta _{+2}^{k}+\Delta _{S}\beta _{-2}^{k}\right] \alpha _{k} &=&0,
\notag \\
\sum_{k}\left[ \Delta _{T}z_{k}\beta _{+1}^{k}+\left( \mu ^{\prime
}+tz_{k}\right) \beta _{+2}^{k}-\Delta _{S}\beta _{-1}^{k}\right] \alpha
_{k} &=&0,  \notag \\
\sum_{k}\left[ -\Delta _{S}\beta _{+2}^{k}+\left( 2B-\mu ^{\prime
}-tz_{k}\right) \beta _{-1}^{k}-\Delta _{T}z_{k}\beta _{-2}^{k}\right]
\alpha _{k} &=&0,  \notag \\
\sum_{k}\left[ \Delta _{S}\beta _{+1}^{k}+\Delta _{T}z_{k}\beta
_{-1}^{k}-\left( 2B-\mu ^{\prime }-tz_{k}\right) \beta _{-2}^{k}\right]
\alpha _{k} &=&0.  \label{alpha}
\end{eqnarray}
Using Eqs. (\ref{zk}) and calling
\begin{equation}
C_{12}=\Delta _{S}\left( \alpha _{1}+\alpha _{2}\right) ,\text{ }
C_{3}=2B-\mu ^{\prime }-tz_{3}-\Delta _{T}z_{3},\text{ }C_{4}=2B-\mu
^{\prime }-tz_{4}+\Delta _{T}z_{4},  \label{cs}
\end{equation}
the last two Eq. (\ref{alpha}) can be written as
\begin{eqnarray}
-C_{12}+C_{3}\alpha _{3}-C_{4}\alpha _{4} &=&0,  \notag \\
C_{12}-C_{3}\alpha _{3}-C_{4}\alpha _{4} &=&0.  \label{ceq}
\end{eqnarray}
The solution of this equation is
\begin{equation}
\alpha _{4}=0\text{, }\alpha _{3}=\frac{C_{12}}{C_{3}},\text{ }C_{3}=B
-\frac{\mu ^{\prime }}{2}+\sqrt{\left( B-\frac{\mu ^{\prime }}{2}\right)
^{2}-t^{2}+\Delta _{T}^{2}},  \label{a34}
\end{equation}
where the expression of $C_{3}$ has been obtained using Eqs. (\ref{zk}) and 
(\ref{cs}). From Eqs. (\ref{zk}), (\ref{cs}), and (\ref{a34}) it is easy to
see that the contribution of $\alpha _{3}$ and $\alpha _{4}$ to the first
two Eqs. (\ref{alpha}) is either of order $\Delta _{S}^{2}$ or zero.
Therefore, it can be neglected to first order in $\Delta _{S}$ leading to
\begin{equation}
\sum_{k=1}^{2}\left( \mu ^{\prime }+tz_{k}+\Delta _{T}z_{k}\right) \alpha
_{k}=0.  \label{aeq}
\end{equation}
Using the expressions for $z_{k}$, the solution can be written in the form 
\begin{equation}
\alpha _{1}=\frac{e^{i\omega }}{\sqrt{2}},\text{ }\alpha _{2}
=\frac{e^{-i\omega }}{\sqrt{2}},\text{ }\omega =\arctan \left[ \frac{\left(
t+2\Delta _{T}\right) \mu ^{\prime }}{2(t+\Delta _{T})c}\right] .
\label{a12}
\end{equation}
Using $|f\rangle =\sum_{k}\alpha _{k}|e_{k}\rangle $,  $|e_{k}\rangle
=\sum_{s i}\beta _{s i}^{k}|js i\rangle $, \ Eqs. (\ref{dtilde}), 
(\ref{map}), (\ref{bloch}), (\ref{zk}), (\ref{a34}), and (\ref{a12}), we
obtain the final expression of the Majorana zero mode at the left end of the
chain (except for a normalization factor)
\begin{equation}
\eta_L =\sum_{j=1}^{N}\left[ \mbox{Re}(e^{i\omega }z_{1}^{j-1})\left( e^{i\pi
/4}d_{j+}^{\dagger }+e^{-i\pi /4}d_{j+}\right) +\frac{\Delta _{S}\cos \omega 
}{C_{3}}z_{3}^{j-1}\left( e^{-i\pi /4}d_{j-}^{\dagger }+e^{i\pi
/4}d_{j-}\right) \right] .  \label{eta}
\end{equation}
The amplitude of the mode is maximum at the first site and decreases
exponentially for sites inside the chain with different decay rates for spin
$+$ and $-$.

In order to make contact to Eqs. (7) and (8) , we need to express $\eta_L$ in terms of the operators $c_{j,\sigma}$ of the original model. To this end, we
introduce the representation of Eqs. (\ref{dac}) in to Eq. (\ref{eta}) and focus on the limit $\lambda \rightarrow 0$. The projection of Eq. (\ref{eta}) on the first site of the lattice
reads $\eta_L= \gamma_L+\gamma_L^{\dagger}$ with 
\begin{equation}
\gamma_L^{\dagger} \sim e^{i \pi/4} \left[c_{1,\uparrow}^{\dagger} + \frac{\Delta_S}{C_3} e^{-i \pi/2} c_{1, \downarrow}^{\dagger} \right].
\end{equation}
We see that this solution has the structure of Eq. (4) with 
\begin{equation}
\delta_L=\pi/4, \;\;\;\;\;\varphi_L=-\pi/2, \;\;\;\;\;\tan(\theta_L/2)= \frac{\Delta_S}{C_3 } 
+O(\frac{ \lambda}{B})
\end{equation}
The results for  $\delta _{L}$ and $\varphi _{L}$ are valid for any value of the parameters in the topological phase with $\Delta,\; t >0$ and $\mu < 0$, with $\vec{n}_{B} \equiv \vec{z}$,
$\vec{n}_{\lambda} \equiv \vec{x}$, and are in full agreement with the result of the continuum model discussed in Section \ref{cont}. The value of $\theta_L$ is however very sensitive to the values
of the parameters of the Hamiltonian. As explained in the main text, our goal is to show that this angle can be inferred from the behavior of the Josephson current in suitably designed junctions. 

In contrast to $\delta _{L}$ and $\varphi _{L}$
(obtained for $\vec{n}_{B}||\vec{z}$ and $\vec{n}_{\lambda }||\vec{x}$), 
$\theta $ depends on the site. As a consequence for other directions of 
$\vec{n}_{B}$ and $\vec{n}_{\lambda }$ (or other systems of coordinates),  
$\delta_{L}$ and $\varphi _{L}$ also depend on the site, since their transformation
properties depend on $\theta $. Nevertheless for the calculation of the
Josephson current we are only interested in the first and the last site of
the chain. 

\section{Numerical calculation of the Josephson current}
\label{numjose}
The Hamiltonian of the system describing two wires and
a Josephson junction is  
\begin{equation}
H(\phi )=H_{\mathrm{w1}}+H_{\mathrm{w2}}+H_{c}(\phi ),
\text{ }H_{\mathrm{c}}=t_{\mathrm{c}}\sum_{\sigma=\uparrow ,\downarrow }\left( e^{i\phi
/2}c_{1R,\sigma}^{\dagger }c_{2L,\sigma}+\text{H.c.}\right) ,  \label{junct}
\end{equation}
where $H_{\mathrm{wi}}$, $i=1,2$, describe two topological superconducting
wires, $\mathrm{w1}$ at the left of $\mathrm{w2}$, described by Eq. (1) of
the main text, and with a difference $\phi =\phi _{1}-\phi _{2}$ between the
superconducting phases, with $\phi =2\pi $ corresponding to one
superconducting flux quantum. The subscript $1R$ ($2L$) indicates the last 
(first)\ site of $\mathrm{w1}$ ($\mathrm{w2}$). Denoting as 
$N_{1}=\sum_{js}c_{1,js}^{\dagger }c_{1,js}$ the operator of total number of
particles of $\mathrm{w1}$, the current flowing through the
junction from left to right is 
\begin{equation}
J(\phi )=\langle e\frac{dN_{L}}{dt} \rangle =\langle \frac{ie}{\hbar }\left[ N_{1},H\right]    \rangle 
=-\frac{et_{\mathrm{c}}}{\hbar }\sum_{\sigma} \mbox{Im}\left[ e^{i \phi/2} \left\langle 
c_{1R\sigma}^{\dagger }c_{2L\sigma}\right\rangle \right] .  \label{j2}
\end{equation}

The above expectation value can be  numerically calculated given the eigenmodes of the Hamiltonian which
correspond to annihilation operators that satisfy   
$\left[ \Gamma _{\nu },H\right] =\lambda _{\nu }\Gamma _{\nu }$, 
with positive $\lambda _{\nu }.$
The relevant part of these operators have the form  
\begin{equation}
\Gamma _{\nu }=\sum_{\sigma}\left[ A_{1R \sigma }^{\nu }c_{1R \sigma}^{\dagger }+A_{2L\sigma}^{\nu
}c_{2L \sigma}^{\dagger }+B_{1R \sigma}^{\nu }c_{1R \sigma }+B_{2L\sigma}^{\nu }c_{2L\sigma}\right] +...,
\label{gamma}
\end{equation}
where ... denotes the contribution of operators at site different from $1R$
and $2L$. 
The coefficients
are known from the numerical diagonalization. Inverting Eq. (\ref{gamma}) we
have
\begin{equation}
c_{1R\sigma}^{\dagger }=\sum_{\nu }\left( \overline{A}_{1R\sigma}^{\nu }\Gamma _{\nu
}+B_{1R\sigma}^{\nu }\Gamma _{\nu }^{\dagger }\right) ,\text{ }c_{2L\sigma}=\sum_{\nu
}\left( A_{2L\sigma}^{\nu }\Gamma _{\nu }^{\dagger }+\overline{B}_{2L\sigma}^{\nu
}\Gamma _{\nu }\right) .\text{ }  \label{cis}
\end{equation}
Replacing in Eq. (\ref{j2}) and taking into account that in the ground state
the only non vanishing expectation values of a product of two $\Gamma _{\nu }
$ and/or $\Gamma _{\nu ^{\prime }}^{\dagger }$ operators is 
$\left\langle \Gamma _{\nu}\Gamma _{\nu }^{\dagger }\right\rangle =1$, 
we obtain
\begin{equation}
J(\protect\phi )=-\frac{et_{\mathrm{c}}}{\hbar }\mbox{Im}\left[ 
e^{i\phi /2}\sum_{\nu \sigma}A_{2L\sigma }^{\nu }
\overline{A}_{1R\sigma}^{\nu }\right].  \label{j3}
\end{equation}

 An alternative expression can be derived from the
numerical derivative with respect of the flux of the eigenvalues  
$\lambda_{\nu }$. This simplifies the diagonalization procedure at 
the cost of
introducing numerical errors in the differentiation.

Noting that only $H_{\mathrm{c}}$ depends on the flux, Eq. (\ref{j2})  can be also related to the ground state energy $E_{g}$  as follows
\begin{equation}
J(\protect\phi )=\frac{d\left\langle H\right\rangle }{d\phi }=\frac{t_{\mathrm{c}}}{2}\sum_{\sigma}
\left\langle ie^{i\phi /2}c_{1R,\sigma}^{\dagger }c_{2L,\sigma}
+\text{H.c.}\right\rangle =\frac{2e}{\hbar }\frac{dE_{g}\left( \phi \right) }{d\phi }.  \label{dhdfi}
\end{equation}
In turn, except for an additive  constant, $E_{g}$ can be calculated as half the sum of all positive eigenvalues of the
Hamiltonian matrix $S=\sum_{\nu }\lambda _{\nu }$. The latter procedure can be justified by using symmetry arguments\cite{alcam} as follows. Considering the
charge conjugation  operation $C$, acting as
$c_{i,j\sigma}^{\dagger }\leftrightarrow c_{i,j\sigma}$ plus complex conjugation. It is
easy to see that $CHC=-H-2\mu N$, where $N=N_{1}+N_{2}$ is the total number
of particles. Taking the number of particles as fixed $\left\langle
N\right\rangle $, we can write this equation in the form 
$\tilde{H}^{\prime }=CH^{\prime }C=-H^{\prime },\text{ }H^{\prime }=H+\mu
\left\langle N\right\rangle$, which can be considered as change of representation of the same
states. Since both $\tilde{H}^{\prime }$ and $H^{\prime }$ have the same 
many-body spectrum but inverted,
the maximum energy of $H^{\prime }$,
which we denote as $E_{M}^{\prime }$ and the ground state $E_{g}^{\prime }$
are related by $E_{M}^{\prime }=-E_{g}^{\prime }$. On the other hand the
state of maximum energy is obtained applying all the creation operators 
$\Gamma _{\nu }^{\dagger }$ to the ground state. Therefore \ $E_{M}^{\prime
}-E_{g}^{\prime }=S=\sum_{\nu }\lambda _{\nu }$, which leads to 
$E_{g}=-\frac{\sum_{\nu }\lambda _{\nu }}{2}-\mu \left\langle N\right\rangle$. Hence,
\begin{equation}
J(\phi )=-\frac{e}{\hbar }\frac{d\sum_{\nu }\lambda _{\nu }}{d\phi }.
\label{j5}
\end{equation}
We have verified that the results of Eq. (\ref{j3}) and (\ref{j5}) coincide
within numerical precision.

\section{Numerical calculation of $\delta_{\nu}$, $\theta_{\nu}$ and $\varphi_{\nu}$}
\label{numajo}

The Majorana modes that enter the effective low-energy Hamiltonian
$H_{\mathrm{eff}}$ for the Josephson current [see Eqs. (12) and (13) of the main text]
have the form
\begin{equation}
\eta _{\nu }=\gamma _{\nu }^{\dagger }+\gamma {_{\nu }},\text{ }\gamma _{\nu
}^{\dagger }=a_{\nu }e^{i\delta _{\nu }}\left[ \cos (\theta _{\nu
}/2)c_{e\uparrow }^{\dagger }+e^{i\varphi _{\nu }}\sin (\theta _{\nu
}/2)c_{e\downarrow }^{\dagger }\right] +...,  \label{eta2}
\end{equation}
where $a_{\nu }$ is a real number that can be chosen positive, the subscript 
$e$ refers to the site at the end of the chain (first or last) where the
Majorana mode is localized and ... refers to the contribution of other sites
which are not important for $H_{\mathrm{eff}}$. The normalization $\eta
_{\nu }^{2}=1$ implies that $a_{\nu }^{2}\leq 1$ is the weight of the end
site in the Majorana mode.  Each fermionic operator $\gamma_{\nu}$ can be expressed as a combination of two Majorana operators $\eta_{\nu}$ and $\tilde{\eta}_{\nu}$ of the form, $\gamma_\nu=\left(\eta_{\nu}+ i \tilde{\eta}_{\nu}\right)/2$, $\gamma_{\nu}^{\dagger}= \left(\eta_{\nu}- i \tilde{\eta}_{\nu}\right)/2$, of which
only $\eta_{\nu}$ contributes at low energy, 
$\gamma _{\nu }^{\dagger }\simeq \eta _{\nu }/2$.

For a finite chain, there is a effective mixing between the Majorana at the
left ($L$) and right ($R$) end of the chain which by hermiticity should be
proportional to $i\eta _{L}\eta _{R}$. Therefore, the one-particle
eigenstates of lowest absolute value correspond to the fermions $f=e^{i\zeta
}(\eta _{L}+i\eta _{R})/2$ and $f^{\dagger }$ which diagonalize $i\eta
_{L}\eta _{R}$. The phase $\zeta $ is unknown. Thus, for the end we are 
interested ($L$ or $R$) we can write, including explicitly only the 
operators related with that end 
\begin{equation}
f=e^{i\zeta^{\prime } }\frac{\eta _{\nu }}{2}+...=Ac_{e\uparrow }^{\dagger
}+Bc_{e\uparrow }+Cc_{e\downarrow }^{\dagger }+Dc_{e\downarrow }+...
\label{fer}
\end{equation}
where the coefficients at the right side are determined by the numerical
calculation. Comparing with Eq. (\ref{eta2}) we see that the parameters of 
$\eta _{\nu }$ can be obtained from the following equations
\begin{equation}
a_{\nu }=2\sqrt{\left( |A|^{2}+|C|^{2}\right) },\text{ }\delta _{\nu }
=\frac{1}{2}\arctan 
\left( \frac{\mbox{Im}\lbrack A/B]}{\mbox{Re}\lbrack A/B]}
\right) ,\text{ }\theta _{\nu }=2\arctan \left( \frac{|C|}{|A|}\right) ,
\text{ }\varphi _{\nu }
=\arctan \left( \frac{\mbox{Im}\lbrack C/A]}{\mbox{Re}\lbrack C/A]}\right) .  \label{param}
\end{equation}

\begin{figure}[h]
\begin{center}
\includegraphics[width=12cm]{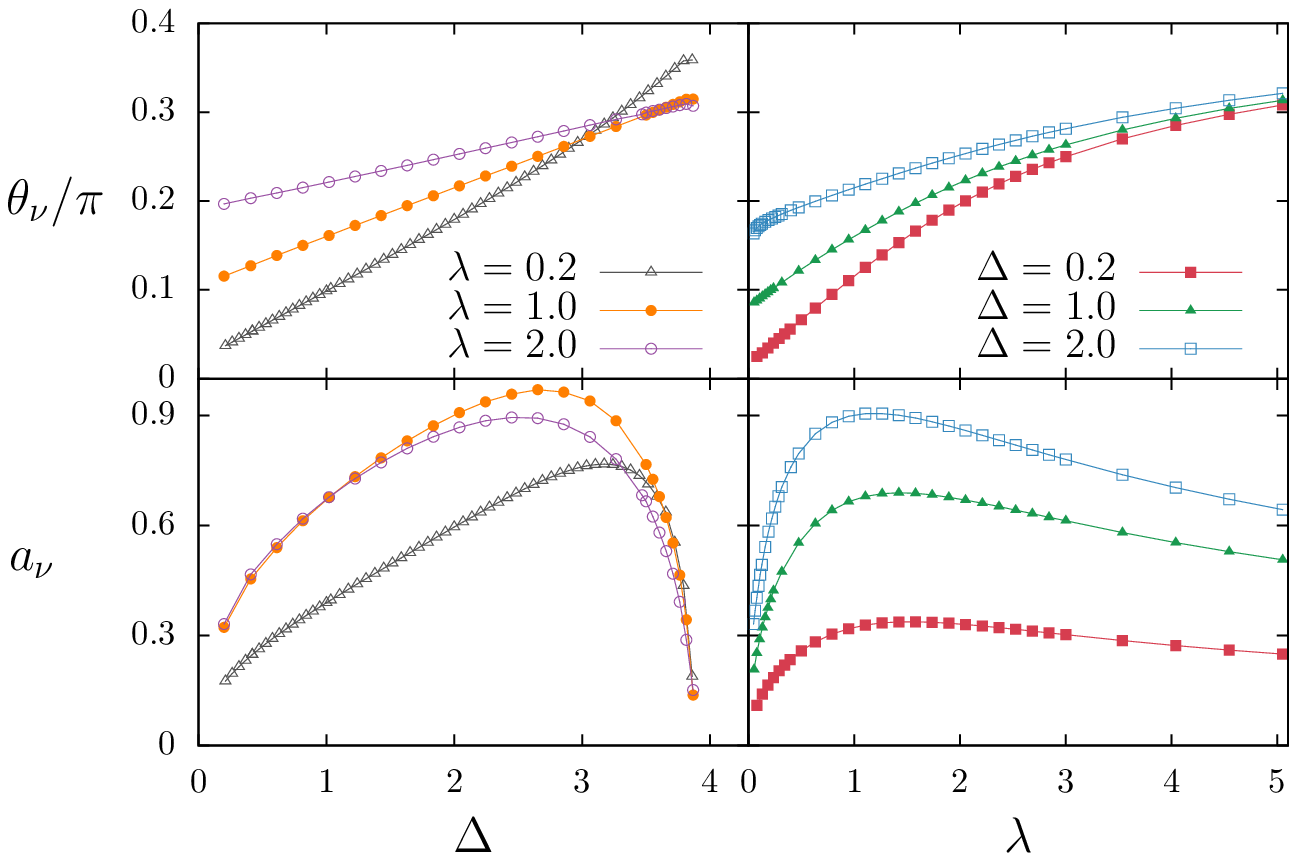}
\end{center}
\caption{Parameters $\theta_\nu$ (top panels) and $a_\nu$ (bottom panels) as a function
of $\Delta$ ($\lambda$) for several  values of $\lambda$ ($\Delta $) and $t=1$, $B=4$
$\mu=-3$, $\vec{n}_B\equiv \vec{z}$ and 
$\vec{n}_{\lambda}\equiv \vec{x}$.}
\label{figpar}
\end{figure}

The dependence of $\theta_\nu$ and $a_\nu$ with the parameters, 
obtained numerically as described above is shown in Fig. \ref{figpar}.
Both determine the coefficient $t_{J}$ of the Josephson current. 
The amplitude $a$ tends to zero at
the borders of the topological region. Curiously, it has a maximum for
intermediate values of $\lambda $.  The angle $\theta $ tends to 0 or $\pi $
(depending on the sign of $\vec{n}_B \cdot \vec{z}$) when both $\lambda $ and $\Delta $
tend to zero as anticipated above.

As explained in the main text, for perpendicular directions of the 
magnetic field and spin-orbit coupling, $\delta_\nu$ and $\varphi_\nu$ can be determined from symmetry arguments and analytical calculations. In particular,
for $\vec{n}_B || \vec{z}$ and $\vec{n}_{\lambda } || \vec{x}$,
\begin{eqnarray}
\delta_L& =&- \delta_R=  \frac{ \pi}{4}, \;\;\;\;\; 
\varphi _L =-\varphi_R= - \frac{\pi}{2}.
\label{delphi}
\end{eqnarray}
In Fig. \ref{delp} we show how these parameters change when the orientation of the spin-orbit coupling $\vec{n}_{\lambda }$ is rotated keeping it in the 
$xy$ plane. We can see that the absolute values of $\delta_\nu$ and 
$\phi_\nu$
increase, keeping $\delta_L = - \delta_R$ and $\varphi _L =-\varphi_R$,
as anticipated in the main text by symmetry arguments.

\begin{figure}[h]
\begin{center}
\includegraphics[width=12cm]{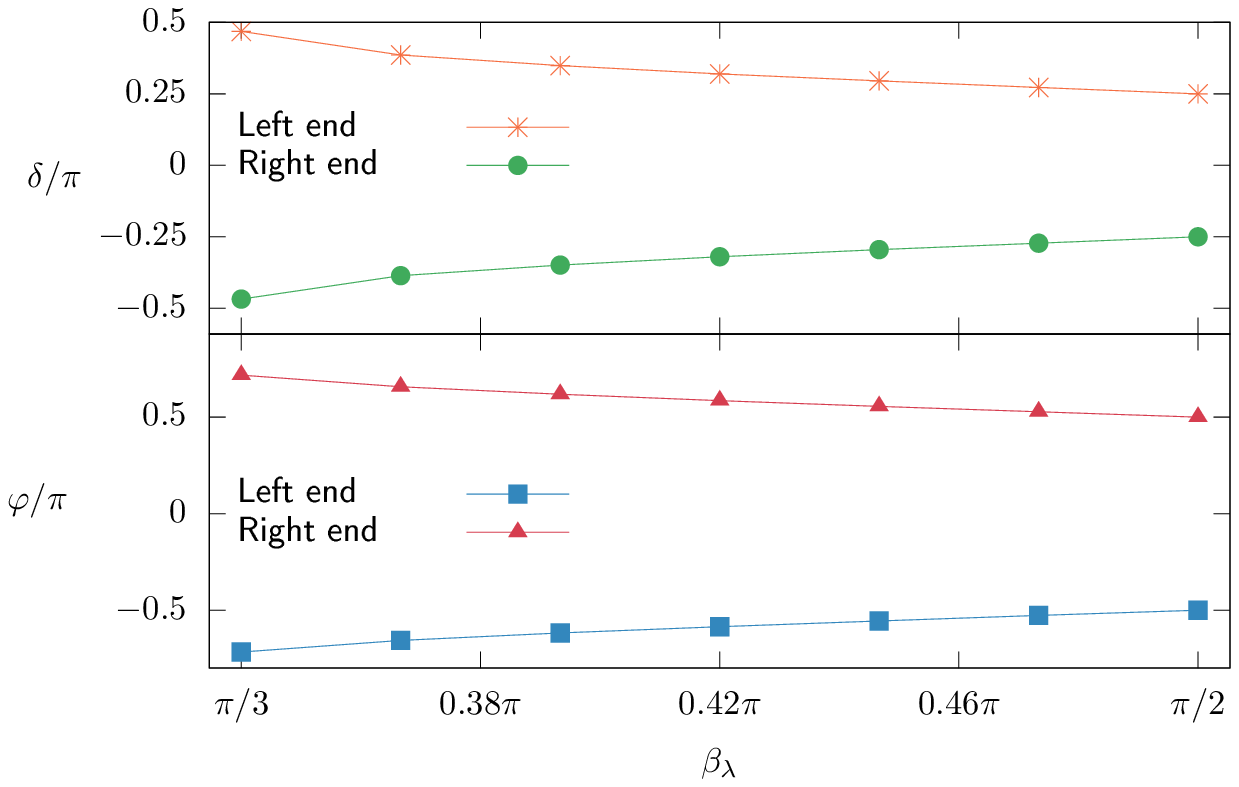}
\end{center}
\caption{Parameters $\delta_\nu$ (top panel) and $\phi_\nu$ (bottom panel) as a function of the angle between the magnetic field and spin-orbit coupling for $t=1$, $B=4$, $\Delta=\lambda=2 $, $\mu=-3$, $\vec{n}_{\lambda}\equiv \vec{x}$, and $\vec{n}_B\equiv \vec{z}$ in the $xz$ plane.}
\label{delp}
\end{figure}

\end{widetext}

\begin{thebibliography}{8}
\bibitem{syma} A change of sign of $\Delta$ can be absorbed by a gauge transformation 
$\tilde{c}_{j \sigma}^{\dagger}=ic_{j \sigma}^{\dagger }$ that has the effect
of flipping the sign of $\delta$ in Eqs. (\ref{apl}). A change of sign of $\lambda$ ($t$)
for $\vec{n}_{\lambda }\cdot \vec{y}=0$ can be compensated  taking the complex conjugate 
of $H$ (complex cojugation and $c_{j \sigma} \rightarrow (-1)^jc_{j \sigma}$ ) and its effect is to change the sign of both $\delta$ and $\varphi$. 
Using time reversal, a change 
in sign of $B$ has the effect on Eq. (\ref{apl}) of replacing $\theta$ by $\pi-\theta$ and inverting
the sign of $\varphi$. Charge conjugation 
$c_{j \sigma}^{\dagger} \rightarrow c_{j \sigma}$
and complex conjugation, which leaves the Majorana invariant, changes the sign of all 
parameters. By consistency a change in sign of $\mu$ 
has the effect of replacing $\theta$ by $\pi-\theta$.
\bibitem{note3} Notice that necessary conditions for a non-vanishing current are breaking of the time-reversal and the spacial inversion symmetry introduced below Eq. (\ref{c1}).
\end{thebibliography}
\end{document}